\documentclass[aps,pre,groupeaddress,showpacs,twocolumn]{revtex4-1}

\usepackage{amsmath}
\usepackage{graphicx}

\usepackage{graphicx}

\begin{document}

\title{Exact density functional for hard rod mixtures derived from
  Markov chain approach}

\author{Benaoumeur Bakhti}
\author{Stephan Schott}
\author{Philipp Maass}
\email{philipp.maass@uni-osnabrueck.de}
\affiliation{Fachbereich Physik, Universit\"at Osnabr\"uck,
Barbarastra{\ss}e 7, 49076 Osnabr\"uck, Germany}

\date{January 19, 2012}

\begin{abstract}

  Using a Markov chain approach we rederive the exact density
  functional for hard rod mixtures on a one-dimensional lattice, which
  forms the basis of the lattice fundamental measure theory.  The
  transition probability in the Markov chain depends on a set of
  occupation numbers, which reflects the property of a
  zero-dimensional cavity to hold at most one particle. For given mean
  occupation numbers (density profile), an exact expression for the
  equilibrium distribution of microstates is obtained, that means an
  expression for the unique external potential that generates the
  density profile in equilibrium.  By considering the rod ends to fall
  onto lattice sites, the mixture is always additive.

\end{abstract}



\maketitle

The extension of density functional theory from continuum to lattice
fluids \cite{Nieswand/etal:1993a} has proven to be useful for treating
problems like ordering transitions
\cite{Nieswand/etal:1993a,Nieswand/etal:1993b,Aranovich/Donohue:2000},
properties of interfaces separating different phases
\cite{Reinel/etal:1994,Prestipino/Giaquinta:2003,Prestipino:2003},
phase separation in mixtures \cite{Woywod/Schoen:2006}, or polymer
adsorption at solid-liquid interfaces \cite{Chen/etal:2009}.
Time-dependent density functional theory \cite{Reinel/Dieterich:1996}
furthermore allows one to describe the kinetics of lattice fluids
\cite{Gouyet/etal:2003}, as emerging in phase ordering phenomena
\cite{Fischer/etal:1998}, relaxation processes
\cite{Heinrichs/etal:2004}, and particle transport in driven lattice
gases
\cite{BramsDwandaru/Schmidt:2007,Dierl/etal:2011,Dierl/etal:2012}.

In 2002 Lafuente and Cuesta extended Rosenfeld's fundamental measure
theory to lattice models based on a derivation of an exact density
functional for hard rod mixtures in one dimension
\cite{Lafuente/Cuesta:2002a,Lafuente/Cuesta:2002b}. This derivation
was carried out following a procedure developed by Vanderlick {\it et
  al.} \cite{Vanderlick/etal:1989} for continuum fluids. Since the
excess free energy part of the functional could be expressed in terms
of differences between parts that agree in their functional form with
the excess free energy functional of a zero-dimensional cavity,
approximate functionals in higher dimensions were obtained by
dimensional expansion of the corresponding difference operator. By
construction these fundamental measure functionals have the property
to become exact under dimensional reduction and their impressive power
was first shown by determining phase diagrams of hard squares
\cite{Lafuente/Cuesta:2002b,Schmidt/etal:2003} and hard cube mixtures
\cite{Lafuente/Cuesta:2002a,Lafuente/Cuesta:2002b,Lafuente/Cuesta:2003a}
with good quality.  The fundamental measure functionals moreover allow
one to apply the method of dimensional crossover and the merit of this
was demonstrated by deriving functionals for lattice gases with
nearest neighbor exclusion for different lattice types (square,
triangular, face- and body-centered cubic) from the functional for
cubes in $(d+1)$ dimensions \cite{Lafuente/Cuesta:2003b}. The
structure of the corresponding results led to a suggestion how to
construct fundamental measure functionals for hard core lattice gases
for any type of lattice, shape of the particles, and arbitrary
dimension \cite{Lafuente/Cuesta:2004}.

In this report we rederive the exact density functional for hard rod
mixtures in one dimension, that means the starting point of the
fundamental measure theory for hard core lattice gases, by applying
the Markov chain approach developed by Buschle {\it et al.}
\cite{Buschle/etal:2000a}. This approach is conceptually different
from the procedure of Vanderlick {\it et al.}
\cite{Vanderlick/etal:1989} and we believe that it is useful and
important on the following reasons: (i) The derivation of the
functional becomes surprisingly simple. Making use only of the
constraints of mutual rod exclusions, the relevant transition
probability in the Markov chain is determined almost without any
calculation. (ii) The transition probability is (conditionally)
dependent on a spatial region, where at most one particle can be
placed, i.e.\ that of a zero-dimensional cavity. In this respect it
reflects a property which turned out to be decisive for the
generalized construction of fundamental measure functionals by
Lafuente and Cuesta \cite{Lafuente/Cuesta:2004}. (iii) The simplicity
of the derivation suggests that it can be extended to hard rod
mixtures with additional (thermal) interactions.  (iv) The derivation
yields also an explicit expression for the probability distribution of
microstates for a given density profile. This means that in the
present case an explicit expression for the ``Mermin potential'' is
obtained, i.e.\ the unique external potential that would generate the
given density profile in thermal equilibrium. In addition to these
points we show that it is not necessary to consider non-additive
mixtures when mixed parities of rod lengths are present (i.e.\ rods
with both even and odd lengths in units of the lattice spacing).

The mixture is considered to consist of $q$ types of hard rods with
length $l_\alpha$, $\alpha=1,\ldots,q$ in the presence of an external
potential.  It is convenient (although not necessary) to order the
lengths according to $l_1\ge l_2\ge\ldots\ge l_q$, where different
types of rods could have the same lengths due to different coupling to
the external potential. The rods are located on a one-dimensional
lattice with $L$ sites and we set the lattice spacing equal to
one. The lattice is defined in such a way that the ends of the rods
coincide with lattice sites and we introduce occupation numbers
$n_j^\alpha$, $j=1,\ldots,L$, $\alpha=1,\ldots,q$, to specify the
microstate of the mixture. If the left end of a rod of type $\alpha$
is at site $j$, then $n_j^\alpha=1$, else $n_j^\alpha=0$ (here and in
the following Greek superscripts refer to the type and must not mixed
up with exponents). The mutual exclusion of hard rods implies the
constraint $n_k^\alpha n_j^\beta=0$ for $j=k,\ldots,k+l_\alpha-1$ (and
$k=j,\ldots,j+l_\beta-1$) \cite{comm:boundary}. In a grand-canonical
description the chemical potentials $\mu_\alpha$ specify the mean
numbers of rods of type $\alpha$.

\begin{figure}[t!]
\centering \includegraphics[width=0.4\textwidth]{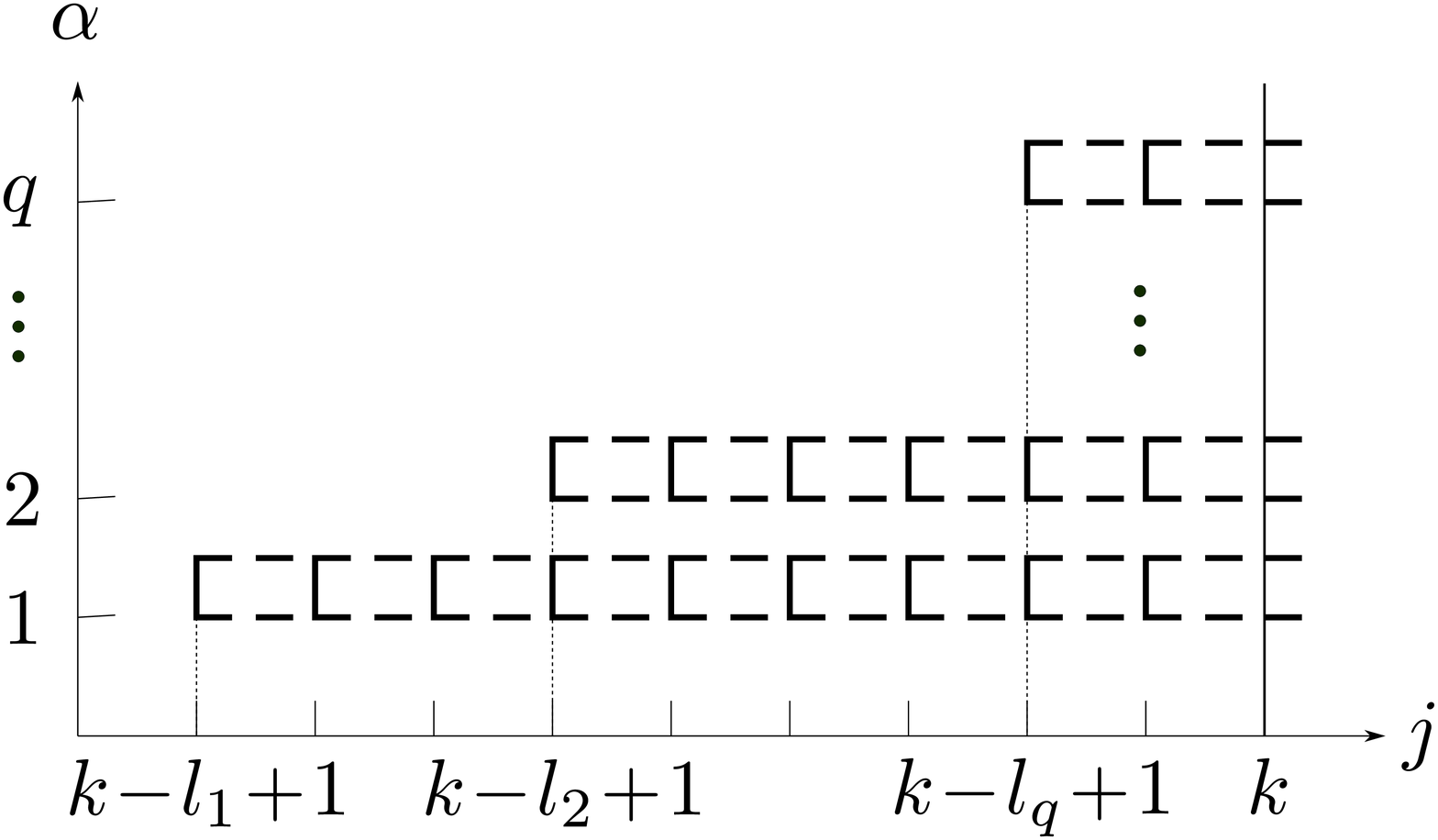}
\caption{Illustration of the set of occupation numbers affecting the
  occupation of site $k$. Any placement of the left end of a rod of
  type $\alpha$ at the sites $j$ with $k-l_\alpha+1\le j\le k$ means
  that site $k$ is covered by a part of this rod. This implies (i)
  that if a left rod end is at site $k$, all occupation numbers in the
  set $\{n_j^\alpha\}_{k-1}=\{n_j^\alpha|1\le\alpha\le q,
  k-l_\alpha+1\le j\le k-1\}$ must be zero, and (ii) that in the set
  $\{n_j^\alpha\}_k=\{n_j^\alpha|1\le\alpha\le q, k-l_\alpha+1\le j\le
  k\}$ there can be at most one occupation number with value 1.}
 \label{fig:fig1}
\end{figure}

To set up the Markov chain approach following
\cite{Buschle/etal:2000a} it is useful to introduce the multicomponent
state variables $\hat n_j=(n_j^1,\ldots,n_j^q)$ that can assume
$(q+1)$ states $\hat e_0,\ldots,\hat e_q$, where $\hat e_0$ refers to
an empty site, i.e.\ $\hat e_0=(0,\ldots,0)$, while $\hat e_\alpha$,
$\alpha=1,\ldots,q$, refer to a site occupied by rods of type
$\alpha$, i.e.\ $\hat e_\alpha=(0,\ldots,1,\ldots,0)$ with the 1 at
the $(\alpha+1)$th entry.  The probability $\chi(\hat n_1,\ldots,\hat
n_L)$ of microstates can be decomposed as
\begin{equation}
\chi(\hat n_1,\ldots,\hat n_L)=\prod_{k=1}^L 
\psi(\hat n_k|\hat n_{k-1},\ldots,\hat n_1)
\label{eq:chi1}
\end{equation}
where $\psi(.|\ldots)$ denote the corresponding conditional
probabilities.  To keep the notation simple, we have labeled the
starting of the chain, i.e.\ $\psi(\hat n_1)\psi(\hat n_2|\hat
n_1)\psi(\hat n_3|\hat n_2,\hat n_1)\ldots$, by the same symbol
``$\psi$'' (meaning in particular that $\psi(\hat n_1)$ is not a
conditional probability). By using the Boltzmann expression for the
probability of microstates in the grand-canonical equilibrium
ensemble, i.e.\ $\chi\propto\exp[-\beta\sum_{i,\alpha}
(u_i^\alpha-\mu_\alpha) n_i^\alpha]$, where $\beta=1/k_{\rm B}T$ is
the inverse thermal energy and $u_i^\alpha$ the external potential, it
can be proven \cite{Buschle/etal:2000b} that the conditional probabilities
satisfy the Markov property
\begin{equation}
\psi(\hat n_k|\hat n_{k-1},\ldots,\hat n_1)
=\psi(\hat n_k|\{n_j^\alpha\}_{k-1})
\label{eq:markov}
\end{equation}
where $\{n_j^\alpha\}_{k-1}=\{n_j^\alpha|1\le\alpha\le q,
k-l_\alpha+1\le j\le k-1\}$ denotes the set of occupation variables,
which have an influence on the occupation of site $k$, see
Fig.~\ref{fig:fig1}.

In the set $\{n_j^\alpha\}_k=\{n_j^\alpha|1\le\alpha\le q,
k-l_\alpha+1\le j\le k\}$, i.e.\ all occupation variables involved in
Fig.~\ref{fig:fig1}, there can be at most one occupation variable
$n_j^\alpha=1$ due to the hard rod constraints, which reflects the
corresponding property of a zero-dimensional cavity.  In fact this set
corresponds exactly to the zero-dimensional cavity for a mixture
introduced in \cite{Lafuente/Cuesta:2002b} as a collection of sets for
each rod type. The property to have at most one occupation
variable $n_j^\alpha=1$ in $\{n_j^\alpha\}_k$ can be utilized to
determine the conditional probabilities by simple probabilistic
considerations. First let us write for $\alpha=0,\ldots,q$
\begin{equation}
\psi(\hat n_k=\hat e_\alpha|\{n_j^\beta\}_{k-1})=
\frac{{\rm Prob}(\hat n_k=\hat e_\alpha,\{n_j^\beta\}_{k-1})}
{{\rm Prob}(\{n_j^\beta\}_{k-1})}
\label{eq:psi-prob}
\end{equation}
where ${\rm Prob}(.)$ denote joint probabilities. If $\alpha\ne0$,
then all $n_j^\beta$ in the set $\{n_j^\beta\}_{k-1}$ must be zero.
This implies ${\rm Prob}(\hat n_k=\hat e_\alpha,\{n_j^\beta\}_{k-1})=
{\rm Prob}(\hat n_k=\hat e_\alpha,\{n_j^\beta=0\}_{k-1})=p_k^\alpha$,
where $p_k^\alpha=\langle n_\alpha\rangle$ is the mean occupation of
site $k$ ($\langle\ldots\rangle$ denotes an average over the
microstate distribution $\chi(\hat n_1,\ldots,\hat n_L)$).  Since with
the same reasoning ${\rm Prob}(\{n_j^\beta=1,\mbox{all other }
n_l^\gamma=0\}_{k-1})=p_j^\beta$, we further have
\begin{equation}
{\rm Prob}(\{n_j^\alpha=0\}_{k-1})+
\sum_{\beta=1}^q\sum_{j=k-l_\beta+1}^{k-1}
p_j^\beta=1
\label{eq:prob-norm}
\end{equation}
due to normalization. Accordingly, we obtain for $\alpha\ne0$
\begin{equation}
\psi(\hat n_k=\hat e_\alpha|\{n_j^\beta\}_{k-1})=
\frac{p_k^\alpha}{1-S_k^{(0)}}
\label{eq:psi-alphane0}
\end{equation}
where we used one of the weighted densities (weighted mean occupations)
\cite{comm:weighting}
\begin{equation}
S_k^{(m)}=\sum_{\alpha=1}^q
\sum_{j=1-m}^{l_\alpha-1}p_{k-j}^\alpha\,,\quad m=0,1
\label{eq:s}
\end{equation}
appearing in the lattice fundamental measure theory
\cite{Lafuente/Cuesta:2002a}.  If $\hat n_k=\hat e_0$ there are two
possibilities: Either one element in $\{n_j^\beta\}_{k-1}$ is one, or
all elements are zero. In the first case, $\hat n_k$ must be equal to
$\hat e_0$, implying that the corresponding conditional probability is
one. In the second case we need ${\rm Prob}(\hat n_k=\hat
e_0,\{n_j^\beta=0\}_{k-1})={\rm Prob}(\{n_j^\beta=0\}_k)$ in
Eq.~(\ref{eq:psi-prob}), which by utilizing normalization as in
Eq.~(\ref{eq:prob-norm}) (now with inclusion of site $k$), is given by
${\rm Prob}(\{n_j^\beta=0\}_k)=
1-\sum_{\beta=1}^q\sum_{j=k-l_\beta+1}^kp_j^\beta=1-S_k^{(1)}$.  In
summary,
\begin{align}
\psi(\hat n_k=\hat e_0|\{n_j^\beta\}_{k-1})&=
\label{eq:psi-e0}\\
&\hspace{-2em}\left\{\begin{array}{cl}
1\,, 
& \mbox{one}\; n_j^\beta=1\; \mbox{in}\; \{n_j^\beta\}_{k-1}\\[2ex]
\displaystyle\frac{1-S_k^{(1)}}{1-S_k^{(0)}}\,, 
& \mbox{all}\; n_j^\beta=0\; \mbox{in}\; \{n_j^\beta\}_{k-1}
\end{array}\right.\nonumber 
\end{align}
Combining Eqs.~(\ref{eq:psi-alphane0}) and (\ref{eq:psi-e0}),
we can write
\begin{align}
\psi(\hat n_k|\{n_j^\beta\}_{k-1})&=
\left(\frac{1-S_k^{(1)}}{1-S_k^{(0)}}\right)^{1-
\sum_{\beta=1}^q\sum_{j=0}^{k-1}n_j^\beta}\nonumber\\
&\hspace{2em}{}\times\prod_{\alpha=1}^q
\left(\frac{p_k^\alpha}{1-S_k^{(0)}}\right)^{n_k^\alpha}
\label{eq:psi-final}
\end{align}
where the distinction between the possible configurations in the
set $\{n_j^\beta\}_k$ is taken into account by the exponents.

Inserting Eq.~(\ref{eq:psi-final}) into Eqs.~(\ref{eq:markov}) and
(\ref{eq:chi1}), the probability distribution of microstates is given
by the product of $\psi(\hat n_k|\{n_j^\beta\}_{k-1})$ from
Eq.~(\ref{eq:psi-final}) over all lattice sites, i.e.\ an explicit
expression for $\chi(\mathbf{n})$ as function of the set
$\mathbf{n}=\{n_i^\alpha|1\le\alpha\le q, 1\le i\le L\}$ of occupation
numbers is obtained (we define $\chi(\mathbf{n})=0$ for all
microstates $\mathbf{n}$ violating the hard rod constraints). This
means that, for a given density profile
$\mathbf{p}=\{p_k^\alpha|1\le\alpha\le q, 1\le k\le L\}$, the
distribution of microstates is uniquely determined if we require it to
satisfy the Markov property (\ref{eq:markov}), i.e.\
$\chi(\mathbf{n})=\chi_{\mathbf{p}}(\mathbf{n})$. One could get the
impression that this is more general than the uniqueness implied by
the Mermin theorem, which states that the prescription of $\mathbf{p}$
fixes the external potential $u_k^\alpha=u_k^\alpha(\mathbf{p})$ in
the sense that the Boltzmann distribution yields $\mathbf{p}$ in
equilibrium in the presence of $u_k^\alpha(\mathbf{p})$. However,
since the Boltzmann distributions satisfy the Markov property
(\ref{eq:markov}), and $\chi_{\mathbf{p}}\mathbf(\mathbf{n})$ is
unique, there is in fact no more generality, i.e.\ the microstate
distribution for given $\mathbf{p}$ satisfying the Markov property
(\ref{eq:markov}) and the Boltzmann distribution generating
$\mathbf{p}$ in equilibrium must be the same \cite{comm:mermin}. We
can thus identify the ``Mermin potential''
$U_{\mathbf{p}}(\mathbf{n})=\sum_{k,\alpha}u_k^\alpha(\mathbf{p})
n_k^\alpha$ by setting $\beta U_{\mathbf{p}}(\mathbf{n})\propto
-\log\chi_{\mathbf{p}}(\mathbf{n})$, which, up to irrelevant constant
contributions, yields (after some rearrangement of summations)
\begin{align}
u_k^\alpha(\mathbf{p})=&\log p_k^\alpha-\log(1-S_k^{(0)})
+
\sum_{j=k}^{k+l_\alpha-1}\log\left(\frac{1-S_j^{(0)}}{1-S_j^{(1)}}\right)
\label{eq:mermin}
\end{align}

Based on the Gibbs-Bogoliubov inequality the density functional in an
external potential $U(\mathbf{n})=\sum_{k,\alpha}u_k^\alpha
n_k^\alpha$ is defined as
\begin{align}
\Omega(\mathbf{p})&=\sum_{\mathbf{n}} \chi_{\mathbf{p}}(\mathbf{n})
\left[k_{\rm B}T\log \chi_{\mathbf{p}}(\mathbf{n})+U(\mathbf{n})-
\sum_{\alpha=1}^q \mu_\alpha N_\alpha\right]\nonumber\\
&=F(\mathbf{p})+
\sum_{k=1}^L\sum_{\alpha=1}^q (u_k^\alpha-\mu_\alpha)p_k^\alpha
\label{eq:omega}
\end{align}
where $F(\mathbf{p})=k_{\rm B}T\sum_{\mathbf{n}}
\chi_{\mathbf{p}}(\mathbf{n})\log \chi_{\mathbf{p}}$ is the free
energy functional.
Inserting $\chi_{\mathbf{p}}(\mathbf{n})$ one obtains
\begin{align}
\beta F(\mathbf{p})=\sum_{k=1}^L&\Bigl\{
(1-S_k^{(1)})\log(1-S_k^{(1)})
\label{eq:f}\\
&{}-(1-S_k^{(0)})\log(1-S_k^{(0)})
+\sum_{\alpha=1}^q p_k^\alpha\log p_k^\alpha
\Bigr\}\nonumber
\end{align}
Minimizing $\Omega(\mathbf{p})$ with respect to the $p_j^\alpha$ yields
the density profile in equilibrium.

Following Lafuente and Cuesta \cite{Lafuente/Cuesta:2002b}, one can
define an ``ideal part'' $F_{\rm id}(\mathbf{p})$ by
\begin{equation}
\beta F_{\rm id}(\mathbf{p})=\sum_{k=1}^L\sum_{\alpha=1}^q
p_k^\alpha(\log p_k^\alpha-1)
\label{eq:fid}
\end{equation}
This differs from the expression $\sum_k\{p_k^\alpha\log p_k^\alpha-
(1\!-\!\sum_\alpha p_k^\alpha)\log(1\!-\!\sum_\alpha p_k^\alpha)\}$
for a non-interacting multi-component Fermionic lattice gas, but has
the advantage to lead to a fundamental measure structure of the excess
free energy part $F_{\rm exc}(\mathbf{p})=F(\mathbf{p})-F_{\rm
  id}(\mathbf{p})$. When using Eqs.~(\ref{eq:f}), (\ref{eq:fid}), and
$\sum_\alpha p_k^\alpha=S_k^{(1)}-S_k^{(0)}$ this becomes
\begin{align}
\beta F_{\rm exc}(\mathbf{p})=
\sum_{k=1}^L&\Bigl\{
\left[S_k^{(1)}+(1-S_k^{(1)})\log(1-S_k^{(1)})\right]\nonumber\\
&\hspace{-2em}{}-\left[S_k^{(0)}+(1-S_k^{(0)})\log(1-S_k^{(0)})\right]
\Bigr\}
\label{eq:fexc}
\end{align}
The terms in the square brackets have the same functional form as the
excess free energy $f_{\rm exc}(\eta)=\eta+(1-\eta)\log(1-\eta)$ of a
zero-dimensional cavity with mean occupation $\eta$
\cite{Rosenfeld/etal:1996}. Approximate fundamental measure
functionals in higher dimensions can be constructed by considering the
two terms in the square brackets as resulting from applying a
one-dimensional difference operator and by generalizing this operator
together with the weighted densities to higher dimensions (for
details, see \cite{Lafuente/Cuesta:2002a,Lafuente/Cuesta:2002b}).

The excess free energy in Eq.~(\ref{eq:fexc}) is equal to that found
by Lafuente and Cuesta for an additive mixture.  To recover their
expressions, occupation numbers $\tilde n_k^\alpha=0,1$ need to
assigned to the rod centers, which amounts to a simple translation of
the site indices, $n_k^\alpha\to \tilde
n_k^\alpha=n_{k+(l_\alpha-\epsilon)/2}^\alpha$, where $\epsilon=0$ if
all $l_\alpha$ are even and $\epsilon=1$ if all $l_\alpha$ are odd.

Non-additive mixtures appear when considering a setup where the rod
centers fall onto lattice sites and both even and odd $l_\alpha$ are
present, since in this case neighboring rods with even and odd
$l_\alpha$ have a minimum separation of half a lattice unit between
their ends.  For such non-additive mixtures one can construct the
corresponding functional from that for additive mixtures
\cite{Lafuente/Cuesta:2002b}. When the rod ends fall onto lattice
sites, the mixtures are always additive irrespective of having mixed
parities of rod lengths.

\begin{acknowledgments}
  We thank J.\ Buschle and W.\ Dieterich for very valuable
  discussions.
\end{acknowledgments}


\begin{thebibliography}{28}
\bibitem [{com({\natexlab{a}})}]{comm:boundary}%
  \BibitemOpen
  \href@noop {} {} \bibinfo {note} {We assume here hard wall
  boundaries, implying that in all expressions $n_k^\alpha=0$ for $k\le0$ and
  $k\ge L-l_\alpha+1$ (same for $p_k^\alpha=\langle
  n_k^\alpha\rangle$).}\BibitemShut {Stop}%
\bibitem [{com({\natexlab{b}})}]{comm:weighting}%
  \BibitemOpen
  \href@noop {} {} \bibinfo {note} {When introducing the
  weighting functions $w_{j,\alpha}^{(m)}$, $m=0,1$, where
  $w_{j,\alpha}^{(m)}=1$ if $j=1-m,\dots,l_\alpha-1$ and zero else, the
  $S_k^{(m)}$ can be expressed by the convolution $S_k^{(m)}=\sum_\alpha\sum_j
  w_{j,\alpha}^{(m)} p^\alpha_{k-j}$.}\BibitemShut {Stop}%
\bibitem [{com({\natexlab{c}})}]{comm:mermin}%
  \BibitemOpen
  \href@noop {} {} \bibinfo {note} {The Mermin theorem does
  not ensure that to each $\mathbf{p}$ a corresponding external potential
  exists (it ensures only that, if it exists, it is unique), but this more
  academic problem seems to be irrelevant for physical
  applications.}\BibitemShut {Stop}%
\end{thebibliography}

%

\end{document}